# Sulfur doping effects on the electronic and geometric structures of graphitic carbon nitride photocatalyst: insights from first principles

Sergey Stolbov and Sebastian Zuluaga

Physics department University of Central Florida, Orlando, FL

**Abstract.** We present here results of our first principles studies of the sulfur doping effects on the electronic and geometric structures of graphitic carbon nitride *(g*-$C_3N_4$*)*. Using the *Ab initio* thermodynamics approach combined with some kinetic analysis, we reveal the favorable S-doping configurations  By analyzing the valence charge densities of the doped and un-doped systems, we find that sulfur partially donates its $p_x$- and $p_y$- electrons to the system with some back donation to the S $p_z$-states. To obtain accurate description of the excited electronic states, we calculate the electronic structure of the systems using the GW method. The band gap width calculated for *g*-$C_3N_4$ is found to be equal to 2.7 eV that is in agreement with experiment. We find the S doping to cause a significant narrowing the gap. Furthermore, the electronic states just above the gap become occupied upon doping that makes the material a conductor. Analysis of the projected local densities of states provides insight into the mechanism underlying such dramatic changes in the electronic structure of *g*-$C_3N_4$ upon the S doping. Based on our results, we propose a possible explanation for the S doping effect on the photo-catalytic properties of *g*-$C_3N_4$ observed in the experiments.

# I. INTRODUCTION

Photocatalytic splitting of water is a promising means for clean production of hydrogen from a renewable source. Various materials have been tested as photocatalysts since Fujishima and Honda [1] first reported the photo-electro-chemical water splitting. There is a number of highly active and stable photocatalysts, such as $TiO_2$ and some other metal oxides. However, due to a wide band gap, they are responsive only to the ultra-violet (UV) range of the spectrum, while the fraction of the UV light in the incoming solar irradiation is only about 4%. Therefore, considerable effort has been made in searching for new stable materials which are photocatalytically active under visible light irradiation [2-5].

It has been recently shown that the graphitic carbon nitrides *(g-$C_3N_4$)* exhibit promising photocatalytic properties [6]. Indeed, these metal free materials have a band gap of 2.7 eV with the $H^+/H_2$ reduction and $O_2/H_2O$ oxidation potentials situated within the gap. Furthermore, the materials are very stable (*g*-$C_3N_4$ is the most stable allotrope of carbon nitrides at ambient conditions [7]) and consist of abundant elements. The pristine *g*-$C_3N_4$ catalyzes the $H_2$ or $O_2$ evolution from water with a sacrificial electron donor or acceptor, respectively, while loading with a small amount of co-catalysts (Pt [8] or Ag [9], for example) enhances hydrogen production rate that is traced to improved carrier separation.

Photocatalytic activity of *g*-$C_3N_4$, however, is still rather low, mostly, for the following reasons: a) the band gap is too wide to efficiently utilize the solar irradiation; b) carrier mobility is restricted due to absence of interlayer hybridization of the electronic states; c) the top of the valence band (VB) is not sufficiently lower than the $O_2/H_2O$ oxidation potential, that may not be favorable for water oxidation [6]. This is thus not surprising that significant effort has been made to improve these properties by changing composition of the materials, in particular, by doping [10]. It has been reported that doping with B [11], P [12], Zn[13], and S [14,15] causes significant changes in optical properties and photocatalytic activity of *g*-$C_3N_4$. For example, the photoreactivity of *g*-$C_3N_4$ toward $H_2$ evolution is found to increase by 7 – 8 times upon sulfur doping [14]. The density functional theory (DFT) [16,17] based calculations performed in that work brought the authors to the conclusion that substitution of the nitrogen atom at the edge of the tri-*s*-triazine units with sulfur is an energetically favorable scenario for the doping. However, the authors analyzed only two possible doping configurations within a simplified total energy consideration. In Ref. 15, sulfur was used only to mediate synthesis of *g*-$C_3N_4$. Nevertheless, the author found that sulfur-containing samples have much higher rates of the $H_2$ and $O_2$ evolution than the pristine *g*-$C_3N_4$. These works show that doping is a promising means for improving photocatalytic activity of *g*-$C_3N_4$. However, it is also clear that rational modification of the properties of *g*-$C_3N_4$ by doping is possible only based on understanding of microscopic mechanisms underlying the doping effects on the properties of interest of *g*-$C_3N_4$. In this work, we make a step toward this understanding by performing systematic computational studies of the electronic and geometric structure of the clean and S doped *g*-$C_3N_4$.

# II. COMPUTATIONAL DETAILS

All calculations have been performed in this work using the VASP5.2.11 code [18] with projector augmented wave potentials [19] and the plane wave expansion for wave functions [20]. The lattice parameters, stacking, structural optimization, energetics of doping and valence electron charge densities have been calculated within the Perdew-Burke-Ernzerhof (PBE) version of the generalized gradient approximation (GGA) for the exchange and correlation functional [21]. The cutoff energy of 750 eV was used for the plane wave expansion of wave functions. To take into account the van der Waals interaction, which essentially determines the interlayer binding in $g$-$C_3N_4$, we added to the DFT part a semi-empirical dispersion potential proposed by Grimme [22].

In order to obtain accurate description of excited electronic states and band gap width we calculated the densities of electronic states of the systems under consideration using the GW method [23], as implemented in VASP5.2.11 [24,25]. The self-energy was evaluated through dynamical screening of interaction with the frequency-dependent dielectric matrix, where the latter was calculated within the random phase approximation. The core – valence electron interaction was treated at the Hartree-Fock level. We found that the one-iteration $G_0W_0$ approximation, in which both Green's function and screened interaction calculated with the self-consistent DFT wave functions, provides the band gap width in agreement with experiment. These results were obtained with the cutoff energy for the response function of 90 eV. We found that increase in the cutoff to 150 eV practically does not change the density of electronic states. The dielectric matrix and Green's function were calculated using 140 bands including 76 unoccupied bands. Increase in the number of bands to 150 did not change results noticeably.

All calculations have been performed for the hexagonal structures with AB staking, for which a $g$-$C_3N_4$ unit cell included two planes populated with 28 atoms ($C_{12}N_{16}$). For such a large unit cell, the (5x5x5) k-point samplings in Brillouin zone used in this work provided sufficient accuracy for the characteristics obtained by integration in the reciprocal space. While performing the structural optimization of the systems, we considered the lattice relaxation to be achieved, as forces acting on atoms did not exceed 0.015 eV/Å.

The geometric structures of the systems and valence charge densities shown in this article were plotted using the Xcrysden software [26].

### III. RESULTS AND DISCUSSION

#### A. Electronic and geometric structures of $g$-$C_3N_4$

First we have obtained the optimized lattice constants and staking of non-doped $g$-$C_3N_4$. The calculated in-plane lattice constant of this hexagonal structure is found to be $a$ = 7.14 Å that is almost exactly equal to the experimental size of the tri-s-triazine unit (7.13 Å [6]), while the calculated $c$ = 6.15 Å is 5.4% smaller than the experimental value [10]. The source of this error is a simple semi-empirical approximation for the van der Waals forces which determine the inter-plane distance. It is important to note, however, that, as we will show, significant changes in $c$ result in a negligible effect on the electronic structure of the system, because the electronic structure of $g$-$C_3N_4$ is determined by a strong in-plane C – N covalent bonding, while the weak inter-plane van der Waals coupling does not cause any noticeable inter-plane hybridization of the electronic states.

To obtain the preferred staking geometry we calculated the total energy for six possible structures with different relative shifts between A and B planes. We have found the structure shown in Fig. 1 to have the lowest energy. In this configuration only unlike atoms (C and N) in different planes are situated against each other (have the same in-plane coordinates) that may lead to a reduced inter-plane electrostatic repulsion. It is worth mentioning that there are other configurations with C and N atoms located against each other and the total energies of these configurations are only by 0.035 – 0.057 eV larger (per unit cell) than the preferred one. Such a small difference in energies suggests that any of these stacking configurations can be found in the synthesized $g$-$C_3N_4$ samples. Meanwhile the structures with C-against-C and N-against-N geometries are found to be less favorable: their total energies are larger than that of the preferred configuration by 0.2 – 0.96 eV per unit cell.

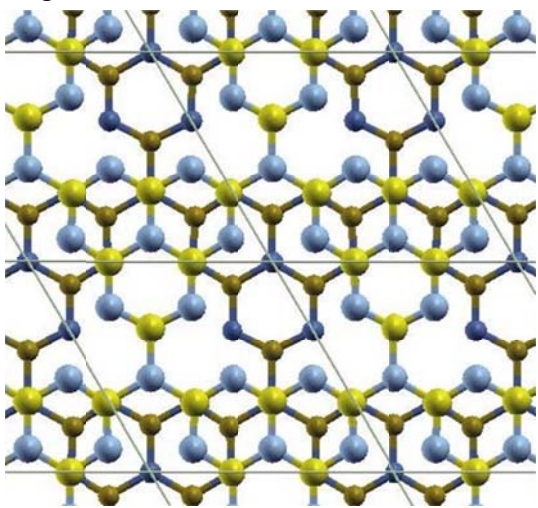

Fig. 1. The lowest energy configuration of the AB stacking obtained for $g$-$C_3N_4$. Yellow and blue balls represent C and N atoms, respectively.

Turning to the electronic structure of $g$-$C_3N_4$, we analyze first the valence charge density $\rho(r)$ in the system. In addition to $\rho(r)$, we calculated difference between the self-consistent valence charge density of the crystal and sum of valence charge densities of isolated atoms placed at the corresponding positions in the crystal:

$$\delta\rho(r) = \rho_{scf}(r) - \sum \rho_{atom}(r) \quad (1)$$

This function describes redistribution of the valence electron density of atoms caused by chemical bonding. Two-dimensional in-plane cuts of both $\rho(r)$ and $\delta\rho(r)$ calculated for $g$-$C_3N_4$ are shown in Fig. 2. The red areas in the left panel of the figure correspond to the electronic charge accumulation along the C – N bonds reflecting a strong covalent bonding between these atoms.

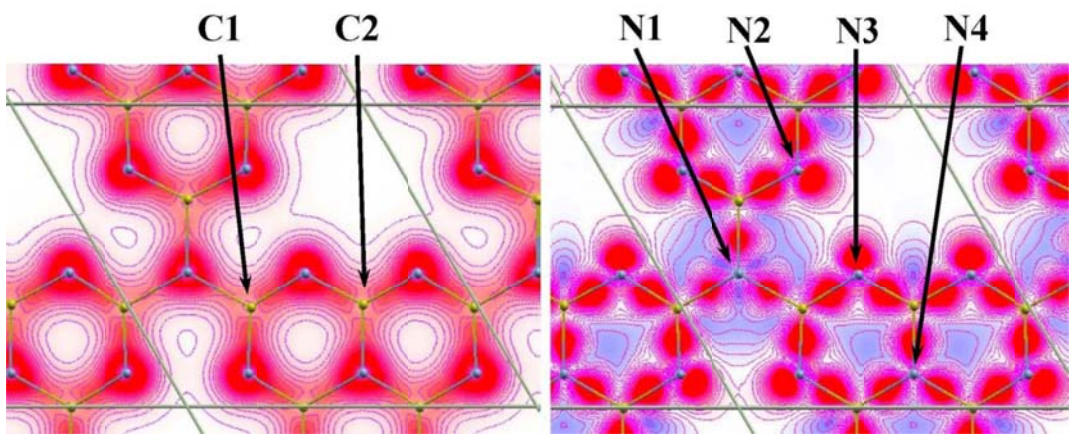

Fig. 2. Left panel: In-plane cut of the valence charge density calculated for $g$-$C_3N_4$. The uniform deep red spots correspond to the regions with $\rho > 0.3$ e/Å$^3$. Right panel: In-plane cut of $\delta\rho(r)$ (see Eq. 1) calculated for $g$-$C_3N_4$. The red and blue colors correspond to the positive and negative values, respectively. The uniform deep red spots mark the regions with $\delta\rho(r) > 0.03$ e/Å$^3$. The lattice site numbering shown in the figure is used in the text.

As expected, the charge redistribution has symmetry typical for the $sp^2$ hybridization. The vertical (xz, yz) $\delta\rho(r)$ cuts (not shown here) suggest that population of the $p_z$-orbitals of both C and N decreases upon chemical bonding. The less intensive red "bridges" between the N1 atom and neighboring C atoms seen in the figure suggest that N1 is bound to the lattice weaker than the other nitrogen atoms. One can also see in the figure two red lobes directed from the atoms N2 and N3 toward the large triangular hollow. This is an important finding, because these lobes represent the nitrogen dangling bonds that make these N sites more reactive than the other sites in the structure. One thus may expect that these lobes make a "trap" for dopant atoms. To test this hypothesis, however, we need to calculate the energetics of doping.

### B. Mechanism of doping of *g*-C$_3$N$_4$ with sulfur

To model the doping of *g*-C$_3$N$_4$ with sulfur, we considered substitutions of N1, N2, N4 and C with S, as well as addition of S to the structure by binding it to the reactive N2 and N3 sites. The calculations were performed for one S atom per unit cell. This corresponds to ~3.5 at% sulfur concentration, which is about five times larger than that reported in the experiment [15]. However, as we will see, the interaction between S atoms and its effect on the electronic structure is still negligible for this concentration. For all calculations, the atomic positions have been optimized upon the doping, while the lattice vectors were kept as for the undoped system considering that the low dopant concentration does not affect the lattice constants noticeably. It is worth mentioning that S is a much larger atom than N or C. Therefore, we found that, for the optimized (relaxed) structures, substitution of N or C with S caused a significant lattice distortion, while, in the case of addition of S to the lattice, the structure kept a planar geometry.

In order to evaluate possible structures of *g*-C$_3$N$_4$ doped with sulfur from point of view of thermodynamics, we calculated the formation energy for all possible configurations of substitution N and C with sulfur, as well as formation energy for the S adsorption to the N dangling bonds revealed in the previous section. The formation energy for the S doping of *g*-C$_3$N$_4$ is defined using a standard methodology:

$$E_{form} = E_{tot}(S{:}C_3N_4) - E_{tot}(C_3N_4) - \mu(S) + \mu(X). \quad (2)$$

Here $E_{tot}(S{:}C_3N_4)$ is the calculated total energy of the doped system, $E_{tot}(C_3N_4)$ is the total energy calculated for pure *g*-C$_3$N$_4$, $\mu(S)$ and $\mu(X)$ denote chemical potential of sulfur and the substituted atom, respectively, X = N or C. We obtain $E_{form}$ for both C-rich and N-rich conditions. For the C-rich condition, C is considered to be in equilibrium with the bulk graphite. Then $\mu(N)$ is determined as

$$\mu(N) + \mu(C) = \mu(C_3N_4), \quad (3)$$

where $\mu(C_3N_4)$ is approximated by the total energy per formula unit. For the N-rich condition, $g$-$C_3N_4$ is assumed to be in equilibrium with gaseous $N_2$ and we can set $\mu(N)=\frac{1}{2}\mu(N_2)$ and $\mu(C)$ is obtained using Eq. 3. The chemical potential of S is calculated for the condition of equilibrium with the rhombic bulk sulfur. We obtain the chemical potentials of C, N, and S for T = 0 K and room temperature (T = 298K) using data on the enthalpy of formation and entropy listed in Ref. [27]. For room temperature, we neglect entropic contribution of $g$-$C_3N_4$ assuming it to be very small (for example, for similar material graphite, TS = 0.06 eV [27]).

Table 1 Formation energies (eV) calculated for different doping configurations

|                | $S_{ad}$ | S ↔ C | S ↔ N1 | S ↔ N2 | S ↔ N4 |
|----------------|----------|-------|--------|--------|--------|
| N-rich, T=0    | 1.690    | 1.546 | 3.059  | 1.520  | 2.231  |
| C-rich, T=0    | 1.690    | 1.998 | 2.720  | 1.181  | 1.892  |
| N-rich, T=298K | 1.292    | 1.272 | 2.568  | 1.029  | 1.740  |
| C-rich, T=298K | 1.292    | 1.997 | 2.023  | 0.484  | 1.195  |

The calculated formation energies are listed in Table 1. Although $E_{form}$ are found to depend on the doping condition, the overall result is that all considered doping scenarios are thermodynamically unfavorable (all numbers in the table are positive). The substitution of the edge N, next to the triangle hollow, is the least unfavorable. It is worth mentioning that the results have been obtained for the highest possible chemical potentials for N and C, which provide the lowest possible formation energies for substitution of N or C by sulfur. Our results thus suggest that the doped system is in a local energy minimum and the formation energies reflect the probability to meet the system at a given configuration in infinite time (if it is in equilibrium with a reservoir that defines the chemical potential of N or C). In this case, kinetics, in particular, the energy barriers to overcome in order to achieve a given configuration become critical factors determining actual result of doping. In this work we study two types of doping: 1) substitution of N or C with S; 2) adsorption of S to the edge of the tri-$s$-triazine sub-units of the crystal, which can be considered as an interstitial doping. It is important to note that there is a significant difference in kinetics of the substitution and adsorption. Indeed, if, for example, $C_3N_4$ is doped with S by heating it in $SH_2$ atmosphere [15], the S-adsorption doping may proceed through the following steps: a) adsorption of $SH_2$ on the active N-site, b) dissociation of the $SH_2$ molecule and binding S to the N2 and N3 sites, c) formation of the $H_2$ in gas phase. In this path only step (b) may have a substantial activation energy barrier. The substitution, in addition to these steps, includes: d) removal of N (or C) atom from the $C_3N_4$ lattice, e) diffusion of S atom to the N (C) vacancy, f) diffusion of N atoms to form $N_2$ (or C atoms to form bulk precipitations). All this steps require overcoming energy barriers and the step (d), according to our calculations has a huge activation barrier (8.41eV) and, therefore, negligible rate. One thus can see that even though substitution of N2 with S is thermodynamically more favorable than addition of S to the N2 and N3 sites, the latter is much more favorable in terms of kinetics and more likely to be achieved in the course of doping. Furthermore, S adsorbed on the N2 and N3 sites has quite high binding energy (1.001 eV) that makes this local energy minimum configuration sufficiently stable. In our further consideration of the S doping effects on the electronic structure of $C_3N_4$ we thus mostly focus on the case of the S-adsorption doping, though we also study the effect of the S ↔ N2 substitution on the band gap of $C_3N_4$.

To understand character of chemical bonding between sulfur and nitrogen atoms in the S doped g-$C_3N_4$, we have calculated the valence charge density of the system. Two-dimensional in-plane cuts of $\rho(r)$ and $\delta\rho(r)$ are plotted in Fig. 3.

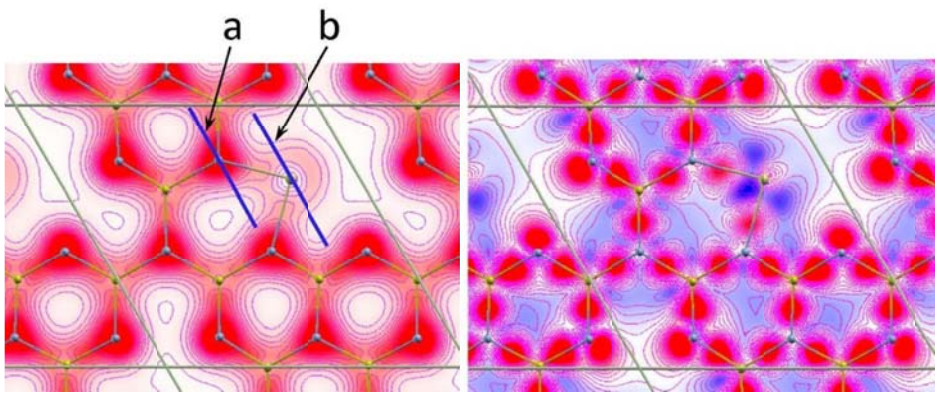

Fig. 3. Left panel: In-plane cut of the valence charge density calculated for g-$C_3N_4$ doped with sulfur. The uniform deep red spots correspond to the regions with $\rho > 0.3$ e/Å$^3$. Right panel: In-plane cut of $\delta\rho(r)$ (see Eq. 1) calculated for g-$C_3N_4$ doped with sulfur. The red and blue collors correspond to the positive and negative values, respectively. The uniform deep red spots mark the regions with $\delta\rho(r) > 0.03$ e/Å$^3$. The sort blue lines mark projection of planes for the $\delta\rho(r)$ cuts shown if Fig. 4.

The figure also shows the preferred position of S in the structure. One can see that a relatively large size of the S atom leads to the elongated S – N bonds (1.75 Å as compared to the 1.32 – 1.42 Å N – C bond lengths). It is also seen in the figure that the valence charge density redistribution in the vicinity of the S – N bonds is significantly different from that in the regions of the C – N bonds. The blue spots around S correspond to the electronic charge depletion that reflects a charge transfer from S $p_x$- and $p_y$-states to the system upon bond formation. Meanwhile, some accumulations of electronic density along the N – S bonds (red spots located closer to N) are also present suggesting complex ionic-covalent character of S – N bonding. These in-plane cuts illustrate the charge density redistribution within the $p_x$- and $p_y$-states of the atoms. Fig. 4 shows the $\delta\rho(r)$ cuts along yz planes that reflect redistribution of the $p_z$-states of S and N neighboring to it. One can see red motifs with a shape of $p_z$-orbitals centered on these atoms showing that occupation of the $p_z$-states of these atoms is increased upon bond formation.

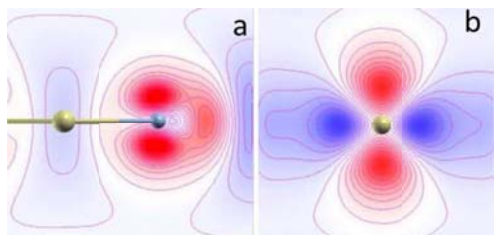

Fig. 4. The cuts of $\delta\rho(r)$ made along the planes whose projections are shown in Fig. 3. The red and blue colors correspond to the positive and negative values, respectively.

## C. Doping effect on the band gap of *g*-C$_3$N$_4$

To reveal a doping effect on the *g*-C$_3$N$_4$ band gap, first we have calculated the densities of electronic states for both un-doped and S-adsorption doped systems using the GW method. It is worth mentioning that changes in stacking configuration, as well as the increase in the interlayer distance by 5.4% practically do not affect the densities of states. This is an expected result, because in these materials there is no noticeable inter-plane overlapping of the electronic states and their electronic structure is totally determined by in-plane hybridization of wave functions.

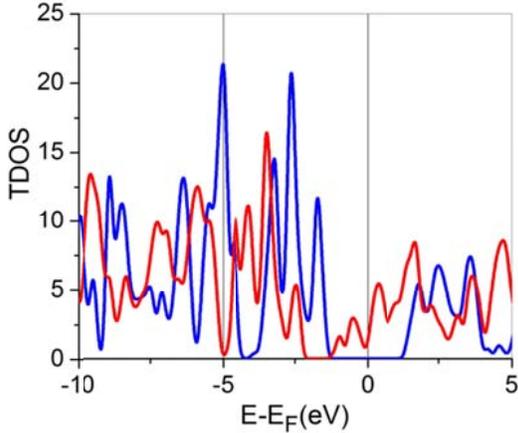

Fig. 5. Total densities of electronic states calculated for pristine *g*-C$_3$N$_4$ (blue line) and *g*-C$_3$N$_4$ with sulfur adsorbed to N2 and N3 atoms (red line) using the GW method.

A part of the total densities of electronic stats (TDOS), plotted for energies in the vicinity of the gap, is shown in Fig. 5. We find the calculated band gap width for *g*-C$_3$N$_4$ to be equal to 2.7 eV that is in agreement with experiment [6]. It is important to mention that the GW methos has already been applied to calculate the electronic structure of various possible phases of C$_3$N$_4$.[28]. Our results are also in agreement with the results of Ref. 28 obtained for the g-C$_3$N$_4$ phase (the band gap width equal to 2.88 eV). We have also calculated the electronic structure of C$_3$N$_4$ within DFT. As expected, these calculations result in highly underestimated band gap width (only about 1 eV). What may call a particular attention in the figure is a dramatic change in TDOS caused by the S doping. One can see that the gap narrows down to ~ 1 eV upon the doping. Furthermore, the Fermi-level is located above the gap suggesting that the S doped *g*-C$_3$N$_4$ becomes a conductor. It is worth mentioning that our DFT calculations show also that the band gap is narrowing upon the doping (from 1 eV to 0.7 eV) and the Fermi-level is shifted to the conducting band. DFT calculations thus show the same trends in the electronic structure modification upon the doping as GW calculations, though the numbers are different. Interestingly, a difference between the band gap widths obtained from GW and DFT calculations is found to be much smaller for the doped system than for the undoped one. On the other hand, it has a simple explanation. Indeed, in the S-doped C$_3$N$_4$ the gap is found to be located between occupied states. This means that only ground states are involved in formation of the gap. Since the ground states are reproduced within DFT well, the DFT and GW redults do not differ much for this case.

These are very important findings, because they may have critical effects on photocatalytic properties of *g*-C$_3$N$_4$. On the other hand, the obtained changes are not surprising. Indeed, electronegativity of S is lower than that of N and one should expect that the extra valence electrons brought with S are donated to the system to occupy a part of conduction band. Our finding is also in agreement with optical absorption measurements [15]. Namely, in Fig. 2 of Ref. [15], one can see long low energy tails in the optical absorption spectra of the sulfur containing $g$-$C_3N_4$ samples, which may correspond to a small fraction of a conducting phase.

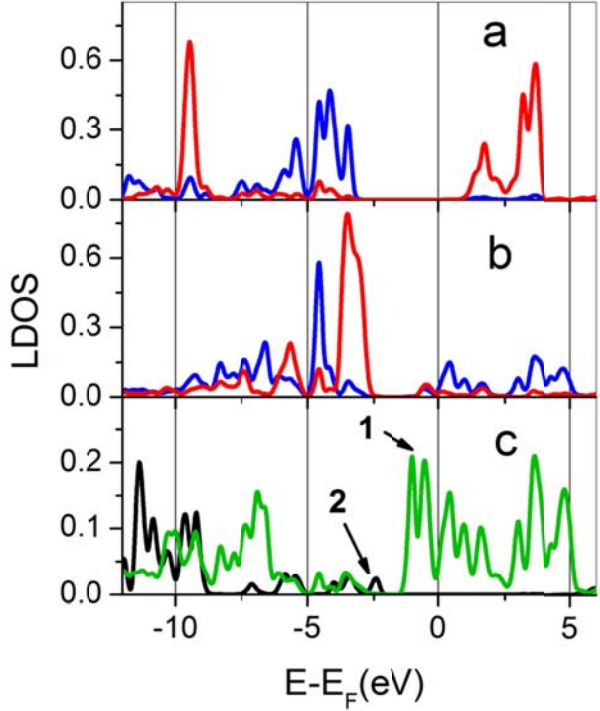

Fig. 6. Projected local densities of states calculated for the S doped $g$-$C_3N_4$ using the GW method. Panel a: N3 $p_y$-states (blue line) and S $p_y$-states (red line). Panel b: N3 $p_z$-states (blue line) and S $p_z$-states (red line). Panel c: C1 $p_z$-states (green line) and C2 $p_y$-states (black line).

To understand the factors controlling the dramatic modification of the electronic structure of $g$-$C_3N_4$ caused by the S doping, we calculated and analyzed the projected local densities of states (LDOS) for all atoms in the unit cell. Some projected LDOS of S, N3 atom neighboring to S, and C1 and C2 atoms neighboring to N3 are plotted in Fig. 6. Due to their symmetry, the S $p_y$- and N3 $p_y$-states considerably overlap that causes a significant energetic splitting of the resulting bands (see plots in the upper panel of the figure). These widely separated bands do not contribute to TDOS within the band gap and thus do not affect its width. Note that a larger fraction of the S $p_y$-states is not occupied, reflecting the fact that S donates its $p_y$-electrons to the system. A similar outcome has been obtained for the $p_x$-states (not shown here). In contrast, the π-symmetry of the S $p_z$- and N3 $p_z$-states determines their weaker overlapping (and hence hybridization), which results in a smaller energetic separation between the corresponding bonding and anti-bonding sub-bands (see panel $b$ in Fig. 6). Consequently, it causes a reduction of the band gap width of the system. We also find that the S doping leads to formation of occupied carbon (C1) $p_z$-states in the upper part of the band gap (a double peak 1 in panel $c$ of Fig. 6). It is critical that these states merge with the conduction band, which makes this material a conductor. An additional narrowing of the gap is caused by formation of a small peak of the $p_y$-states of the C2 atom (peak 2 in panel $c$ of Fig. 6) just above the valence band. Some nitrogen atoms also contribute to TDOS in this region.

Although our analysis suggests that S adsorption to N2 and N3 is most likely to be achieved in the course of doping, we also calculate the electronic structure for the other possible configuration, namely for the least unfavorable substitution: S ↔ N2. TDOS, calculated for this configuration within the GW method is plotted for energies in the vicinity of the gap in Fig. 7.

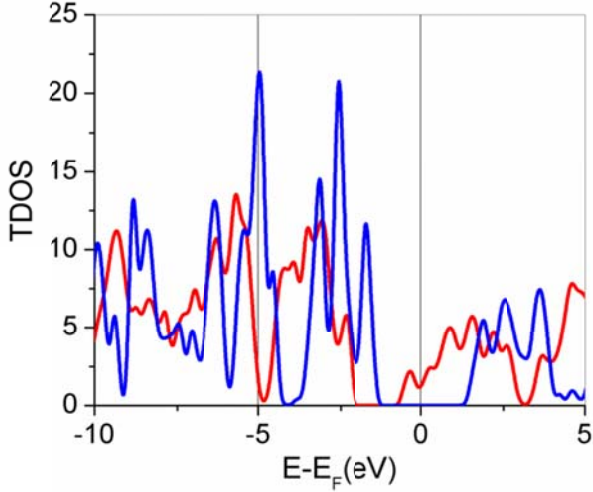

Fig. 7. Total densities of electronic states calculated for pristine $g$-$C_3N_4$ (blue line) and $g$-$C_3N_4$ with sulfur S substituting the N2 atom (red line) using the GW method.

As seen from the figure, substitution of N2 with S also causes narrowing of the gap and shifts the Fermi-level to the conduction band.

Our results have thus brought us to the conclusion that sulfur doping makes $g$-$C_3N_4$ a conductor. Such transformation is very unfavorable for a photocatalyst. On the other hand, the authors of Ref. [14,15] report an enhancement of photocatalytic activity of $g$-$C_3N_4$ upon the S doping. We can explain this "discrepancy" based on the assumption that sulfur is not uniformly distributed over the $g$-$C_3N_4$ studied in Ref. [14,15]. In this case, the un-doped fractions of the material work as a photo-anode, while the doped conducting inclusions serve as co-catalysts collecting electrons. Such a modification of $g$-$C_3N_4$ would improve charge separation and thus enhance the hydrogen evolution rate.

IV. CONCLUSIONS

First principle calculations of the electronic and geometric structure of graphitic $C_3N_4$ and that doped with sulfur have been carried out. The DFT total energies calculated for $g$-$C_3N_4$ with different stacking revealed the most favorable structures of this layered crystal. Based on the combined thermodynamics and kinetics analysis we found that substitutions of C or N with S are not favorable. The S atoms rather prefer to adsorb to reactive N sites at the edges of the tri-$s$-triazine units. The valence charge densities obtained for both doped and un-doped $g$-$C_3N_4$ suggest that S donates electronic charge from its $p_x$- and $p_y$-state to the system, while some back donation to the S $p_z$-states also takes place. The band gap width obtained from our GW calculations is in agreement with experiment. We find the S doping to cause dramatic changes in the electronic structure of $g$-$C_3N_4$. It leads, in particular, to a significant narrowing of the band gap and shift of the Fermi-level to the conduction band that makes the material a conductor. We assume that, in the samples studied experimentally in Ref. [14,15], sulfur is non-uniformly distributed over the material. Then, un-doped fractions of $g$-$C_3N_4$ work as photo-anode, while doped inclusions collect excited electrons. As a result, charge separation is promoted and photo-catalytic activity of the material is increased.

**Acknowledgement** We thank Zachary Williams for help with calculations of the staking energies of $g$-$C_3N_4$.